\documentclass[10pt,twocolumn,letterpaper]{article}
 \usepackage{cvpr}              
\usepackage{xcolor}
\usepackage{multirow}
\usepackage{adjustbox}
\usepackage{csquotes}
\usepackage[accsupp]{axessibility}
\usepackage{amsmath}

\usepackage{fancyhdr}

\definecolor{cvprblue}{rgb}{0.21,0.49,0.74}
\usepackage[pagebackref,breaklinks,colorlinks,citecolor=cvprblue]{hyperref}

\def\paperID{13226} 
\def\confName{CVPR}
\def\confYear{2024}

\title{Universal Robustness via Median Randomized Smoothing for Real-World Super-Resolution} 

\author{Zakariya Chaouai\\
{\tt\small zakariya.chaouai@cea.fr}
\and
Mohamed Tamaazousti\\
{\tt\small mohamed.tamaazousti@cea.fr}
\\
\hspace{-5.8cm}{Université Paris-Saclay, CEA, List, F-91120, Palaiseau, France}    
}

\begin{document}
\maketitle
\newtheorem{lemma}{Lemma}[section]
\begin{abstract}

Most of the recent literature on image Super-Resolution (SR) can be classified into two main approaches. The first one involves learning a corruption model tailored to a specific dataset, aiming to mimic the noise and corruption in low-resolution images, such as sensor noise. However, this approach is data-specific, tends to lack adaptability, and its accuracy diminishes when faced with unseen types of image corruptions. A second and more recent approach, referred to as Robust Super-Resolution (RSR), proposes to improve real-world SR by harnessing the generalization capabilities of a model by making it robust to adversarial attacks. To delve further into this second approach, our paper explores the universality of various methods for enhancing the robustness of deep learning SR models. In other words, we inquire: \enquote{Which robustness method exhibits the highest degree of adaptability when dealing with a wide range of adversarial attacks ?}. Our extensive experimentation on both synthetic and real-world images empirically demonstrates that median randomized smoothing (MRS) is more general in terms of robustness compared to adversarial learning techniques, which tend to focus on specific types of attacks. Furthermore, as expected, we also illustrate that the proposed universal robust method enables the SR model to handle standard corruptions more effectively, such as blur and Gaussian noise, and notably, corruptions naturally present in real-world images. These results support the significance of shifting the paradigm in the development of real-world SR methods towards RSR, especially via MRS.
\end{abstract}     
\section{Introduction}
\label{sec:intro}

The aim of single-image super-resolution (SISR) is to improve the resolution of a given low-resolution (LR) image, by producing a high-resolution (HR) image that is clear and without artifacts. SISR is widely used in a range of real-world applications, such as oceanography \cite{Ducournau}, surveillance \cite{ZhangLiangpei}, and medical images \cite{HuangYawen}. However, super-resolving an image poses a considerable challenge due to the ill-posed nature of the problem, since multiple HR solutions can correspond to a single LR image. There are several well-known methods for scaling high-resolution images, such as linear interpolation methods \cite{KeysCubicconvolutioninterpolation} or the estimation of covariance or correlation in LR data \cite{AllebachEdge-directedinterpolation, LINewedge-directedinterpolation}. Unfortunately, these methods often produce results that appear blurred, noisy and have difficulty in faithfully capturing high-frequency image details.
 
In recent years, SISR methods based on deep neural networks (DNNs) have made considerable progress \cite{Ledig1, LimBee, ShocherAssaf, DongChao, ZhangWenlong} and offer much better quality for the upscaled image. Despite this progress, DNNs have been shown to be vulnerable to adversarial attacks, whether in classification \cite{SzegedyIntriguing, GoodfellowExplaining, MardyTowards} or in SR \cite{ChoiJun-Ho1, ChoiJun-Ho2} (see Figure \ref{fig:Attack_illustration}). The inevitability and universality of adversarial examples is rooted in their definition. It is possible to systematically introduce additive perturbations into the input, causing the model to misclassify an example. The susceptibility to adversarial inputs poses a potential issue, hindering the application of deep learning methods in security and safety-critical contexts. It is important to note that even state-of-the-art SR models \cite{Lim2017, ZhangWenlong} tend to perform poorly on real-world images that contain some corruption or amount of sensor noise. Since the majority of SR models are trained in a supervised way, requiring matching pairs of HR and LR images, LR images are typically generated from HR images by using bicubic downscaling. 

The recognition of this constraint spurred the investigation of real-world SR on datasets with synthetic and natural corruptions. Several benchmarks \cite{LugmayrAndreasNtire2020challenge, LugmayrAndreasAim2019challenge} design real-world artifacts and corruptions under different assumptions or from varying sensors. Consequently, some methods in real-world SR \cite{FritscheManuelseparationforreal-worldsuper-resolution, JiReal-worldsuper-resolutionviakernelestimationandnoiseinjection} generate photo-realistic results only when they are evaluated on a specific dataset for which they were trained, but they fail to generalize to new datasets with unseen corruptions. A more recent approach, Castillo et al. \cite{CastilloAngela}, referred to as Robust Super-Resolution (RSR), proposes to improve real-world SR by harnessing the generalization capabilities of a model, making it robust to unseen noise by using adversarial training, see Subsection \ref{Adv-train}. To the best of our knowledge, it is the only work that has attempted to create a generalized real-world SR model that achieves state-of-the-art results without training or fine-tuning on real-world datasets.

In this paper, we delve further into this latter approach. We recall that the adversarial learning employed in \cite{CastilloAngela} relies on using the Projected Gradient Descent (PGD) attack \ref{PGDAttack} as a form of attack on LR images during the training phase. However, we will show that this type of defense is sensitive to other types of perturbations, and it is not the most effective generalized real-world SR model. In response to this limitation, we employ the Median Randomized Smoothing (MRS) approach, a scalable technique providing certified robustness for neural network-based models. This technique, initially applied in the context of object detection \cite{chiang2020detection}, transforms any DNN into a new smoothed one with certifiable $l_2$-norm robustness guarantees, as described in Lemma \ref{certificationlemma}. The transformation is defined as follows: let $f_{\theta}: [0,1]^{n}\to [0,1]^{m}$, $f_{\theta} = (f^{1}_{\theta}, ..., f^{m}_{\theta})$,  be a SR neural network, and $x$ be an input. Then, the median smoothing of $f_{\theta}$ is defined as $q_{0.5}(x)= (q^{1}_{0.5}(x),..., q^{m}_{0.5}(x))$, where  $q^{i}_{0.5}(x) = \inf\{y\in \mathbb{R} | \mathbb{P}(f^{i}_{\theta}(x+G)\leq y)\leq 0.5\}$ and $G\sim N(0,\sigma^{2} I)$ follows a Gaussian distribution. The estimation of $q_{0.5}(x)$ can be approximated empirically through Monte Carlo (MC) sampling, as explained in \cite{chiang2020detection}. The advantage of using the median on SR over the mean, commonly used in the classification field \cite{salman2019provably}, stems from the fact that the median is nearly unaffected by outliers present in LR images. Unlike the median, the mean tends to smooth out the areas where predictions are locally constant, \cite{chiang2020detection}, which is disadvantageous for images as they often contain textures. Moreover, it is important to mention that the MRS method is known to require a large number of samples (of order 2000 \cite{chiang2020detection}) with the MC procedure for classification and regression in object detection tasks. However, we discovered that in the context of SR, the MSR is well-suited because pixel-wise variations in predicted images are not large. We can easily control this instability with a few samples (of order 21). Finally, we will need to fine-tune the SR model on noisy LR images using different Gaussians samples to make it insensitive to this type of noise, as we are certifying our model with this type of noise, for our SR model to be insensitive to this type of noise. 

\noindent Our main contributions are as follows:
\begin{enumerate}
    \item We extend the use of adversarial attacks in  SR. Until now, only the PGD attack presented in \cite{CastilloAngela} has been applied in the context of real-world SR based on the perceptual loss. In this paper, we adapt other commonly used attacks from the classification literature.  Specifically, we adapt the Fast Gradient Sign Method (FGSM), the Basic Iteration Method (BIM), and the Carlini and Wagner (CW) attack to the perceptual and pixel level of the image. We apply adversarial training using these attacks to create RSR models.
    \item We propose a novel use of MRS to create a real-world SR model named CertSR that achieves state-of-the-art results, particularly for the Learned Perceptual Image Patch Similarity (LPIPS) metric. 
    \item Finally, we show that MRS is more universal in terms of robustness compared to all the previously mentioned adversarial training techniques.
\end{enumerate}

\section{Related works}
It has been shown by Choi et al. \cite{ChoiJun-Ho1} that state-of-the-art deep learning-based SR methods are highly susceptible to adversarial attacks. This vulnerability is primarily attributed to the propagation of the perturbation through the convolutional operation. In the SR domain, adversarial examples can be represented as follows: an original LR image  $x$ is perturbed by adding a small value $\delta$ to generate an adversarial LR image $x_{adv}$. Consequently, $x_{adv}$ is slightly different from $x$. However, the prediction of  $x_{adv}$ deteriorates significantly compared to the prediction of $x$.
 
 We note that adversarial attacks and robust models are applied to SR for the first time by Choi et al. \cite{ChoiJun-Ho1, ChoiJun-Ho2}. Notably, Choi et al. \cite{ChoiJun-Ho1} explored target and non-targeted attacks, originally developed for classification tasks by Kurakin et al. \cite{KurakinAdversarial}. They adapted these attacks to SR with the goal of maximizing the pixel degradation of super-resolved images. In \cite{ChoiJun-Ho2}, Choi et al. proposed a defense method formulated as an entropy regularization loss for model training, against the adversarial attacks constructed in \cite{ChoiJun-Ho1}, thus improving the robustness of the original SR model. However, as explained in \cite{CastilloAngela},  these last works focused on evaluating the methods based on pixel-wise metrics and did not concentrate their study on real-world SR.

It is worth mentioning that in the context of SR, the primary objective is to obtain perceptually well-resolved HR images. In pursuit of this objective, Castillo et al. \cite{CastilloAngela} recently employed an adversarial attack based on pixel-wise and perceptual losses to construct a robust model. This type of attack was originally introduced by Madry et al. \cite{MardyTowards} for classification tasks. To the best of our knowledge, this work is the only one that reports the study of adversarial training for real-world SR problems, where the evaluation was done on perceptual metrics. In this study, we will show that our method performs much better, is more robust, and generalizes better to real-world SR problems, achieving state-of-the-art results without training or fine-tuning on corrupt datasets.

\section{Adversarial attacks and training on SR}\label{AdvAttTrain}
\begin{figure}[htp!]  
  \centering 
  \includegraphics[width=\linewidth]{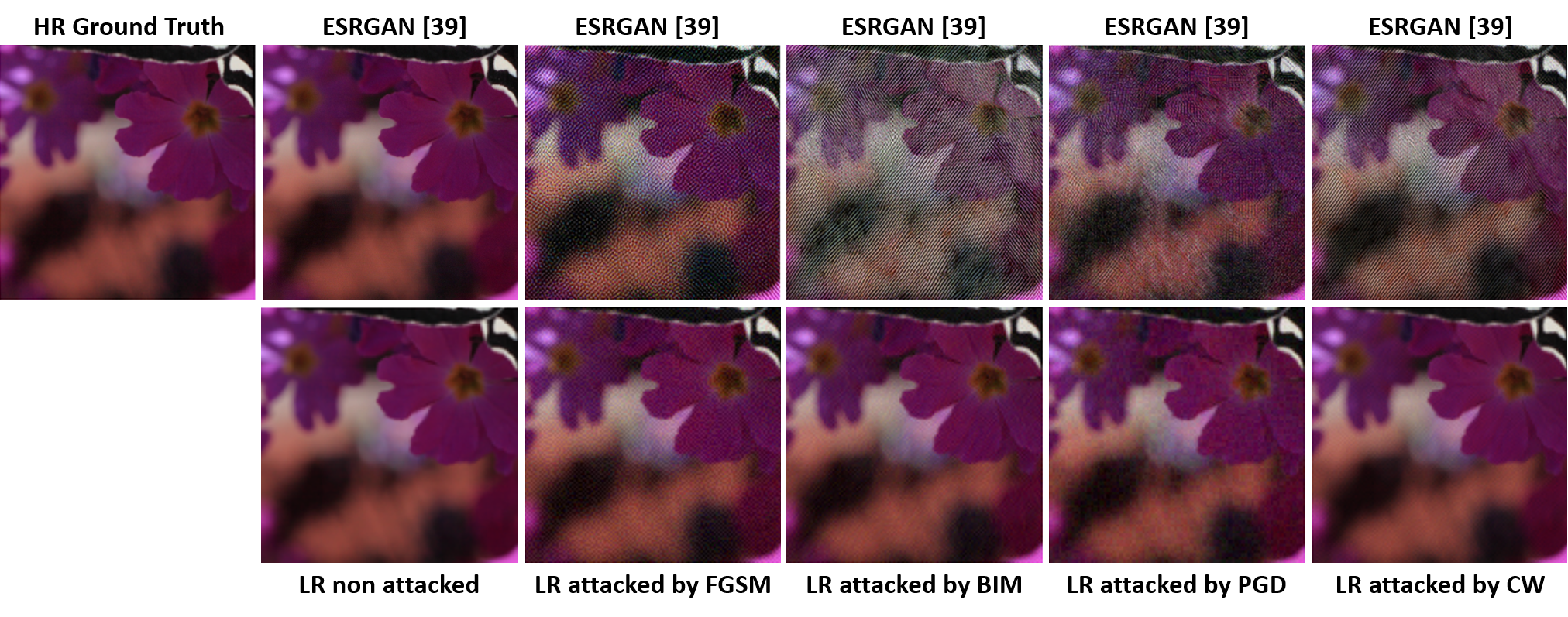}
  \caption{Visualization of both non-attacked and the corresponding attacked LR image subjected to various types of attacks, which we presented above, along with their predictions using ESRGAN \cite{LimBee}, is provided in the first row. The top-left corner displays the ground truth image from the validation dataset of DIV2K \cite{AgustssonNtire2017challenge}, while the clean LR image is shown below it. The LR image was attacked using FGSM, BIM, and PGD with perturbations bounded within a ball of radius $\epsilon=10/255$. For the CW attack, we utilized Adam \cite{Adam} optimization to solve the problem in \eqref{CW_Optimization} with a learning rate of $10^{-2}$ for 6 iterations and $c=0.01$.} 
  \label{fig:Attack_illustration}
\end{figure}

In this section, we present novel adversarial attacks tailored for SR tasks. It is noteworthy that these attacks are drawn from the most relevant and widely employed techniques in the classification literature \cite{GoodfellowExplaining, KurakinAdversarial, MardyTowards, CarliniNicholas}. The visual effect of these adversarial attacks is revealed in Figure \ref{fig:Attack_illustration}. Subsequently, we will provide a general overview of adversarial learning, regardless of the specific adversarial attack used. These adversarial attacks, as well as the RSR based on these attacks, will be used in our experiments to assess the universality of the robustness of our certified SR approach. This evaluation encompasses various adversarial attacks, perturbations existing in the literature, and synthetic perturbations representative of those encountered in real-world images (as detailed in Section \ref{Experiments}). 

\subsection{Adversarial attacks}
\paragraph{Fast Gradient Sign Method (FGSM)}\label{FGSMAttack} is primarily designed to be a fast algorithm for generating adversarial LR images. Moreover, it is an attack that uses the gradient of the loss function to determine the direction in which pixel intensities should be changed to find the most efficient input perturbation. The adversarial LR image is mathematically calculated as follows:
\begin{equation}
    x_{adv} = x +\epsilon\text{ sign}(\nabla_x \mathcal{L}(f_{\theta}(x), y)),
    \label{eq:FGSM}
\end{equation}
where $\mathcal{L}$ is composed of the $L_{percep}$ perceptual and $L_{1}$ pixel-wise loss functions of the generator. Here, $x$ represents the LR image, $y$ represents the HR ground truth, and $\epsilon$ is the step size for the allowed perturbation. As $\epsilon$ increases, it becomes easier to degrade the network's predictions. 

\paragraph{Basic Iterations Method (BIM)}\label{BIMAttack} represents a simple
refinement of the FGSM attack. Instead of taking a single step of size $\epsilon$ in the direction of the gradient sign, multiple smaller steps $\alpha$ are taken. Specifically, begin by setting $x_{0} = x$
as a clean LR image used for initialization in iteration,
\begin{equation*}
    x_{t} =x_{t-1}+ \alpha \text{ sign}(\nabla_{x_{t-1}} \mathcal{L}(f_{\theta}(x_{t-1}), y)).
\end{equation*}
Here, $\alpha=\frac{\epsilon}{T}$, where $T$ represents the number of iterations. This approach is convenient because it provides extra control over the attack. 

\paragraph{Projected Gradient Descent (PGD) \cite{CastilloAngela}}\label{PGDAttack} is considered as a generalization of the BIM attack that doesn't require the condition $\alpha=\frac{\epsilon}{T}$. Moreover, the initialization begins with perturbed LR images following a uniform distribution $U(-\epsilon, \epsilon)$. The perturbation is computed by taking multiple steps of gradient ascent with a small step size $\alpha$ and then projecting the perturbation onto the $\epsilon$-ball around the input. Specifically, start by setting $ x_{0} = x + u$, where $x$ is a clean LR image and $u\sim U(-\epsilon, \epsilon)$ is used for initialization in iteration,
\begin{equation*}
    x_{t} =\text{clip}_{x,\epsilon}(x_{t-1}+ \alpha \text{ sign}(\nabla_{x_{t-1}} \mathcal{L}(f_{\theta}(x_{t-1}), y))).
\end{equation*}
Here, $\text{clip}_{x,\epsilon}$ denotes the clipping of the values of the adversarial sample so that they fall within an $\epsilon$-neighborhood of the original sample $x$.

\paragraph{Carlini and Wagner attack (CW)}\label{CWAttack} is an optimization-based adversarial attack. In this attack, the perturbation is not constrained by the $\epsilon$-ball in the infinite norm but aims to be minimal for the $L_{2}$ norm. The goal of this attack is to maximize the loss function by attacking images with the optimal perturbation. The optimization problem is given by:
\begin{equation}\label{CW_Optimization}
\min_{\delta}(\Vert\delta \Vert_{2} -c\cdot\mathcal{ L}(f_{\theta}(x), y)), \text{ such that } x+\delta \in [0,1]^{n},  
\end{equation}
where $c$ is a hyperparameter. To ensure that $x+\delta \in [0,1]^{n}$, which means that $x+\delta$ yields a valid image, it introduces a new variable $w$ to substitute as follows
$$\delta = \frac{1}{2}(\tanh(w)+1)-x.$$
\subsection{Adversarial training}\label{Adv-train}
Roughly speaking, adversarial training consists of using adversarial examples generated from the training data set to increase robustness locally around the training samples. In this paper, in addition to our main method, which will be presented in Section \ref{mainmethod}, we will employ this technique to create robust models for comparison. 

Adversarial learning typically takes the form of a robust min-max optimization problem, that is given as follows,
\begin{equation*}
    \theta^{*}_{\mathrm{adv}} =\text{\small{argmin}}_{\theta\in\Theta}\dfrac{1}{N} \sum_{(x^{(i)},y^{(i)})\in\mathcal{D}}\max_{\Vert\delta\Vert_{2}\leq\epsilon}\mathcal{L}(f_{\theta}(x^{(i)} + \delta),y^{(i)}),
\end{equation*}
where $\mathcal{D}$ is a batch of LR and HR images. The training is usually processed using an optimization algorithm based on gradient descent on mini-batches. It is important to note that at each iteration of the optimization process, the DNN parameters are updated, and it is necessary to compute the adversarial perturbations with respect to these new parameters at each iteration. This step requires a huge additional computation time compared to classical learning.

\section{The Main Method}\label{mainmethod}

\begin{figure*}[htp!]
  \centering
  \includegraphics[width=\linewidth]{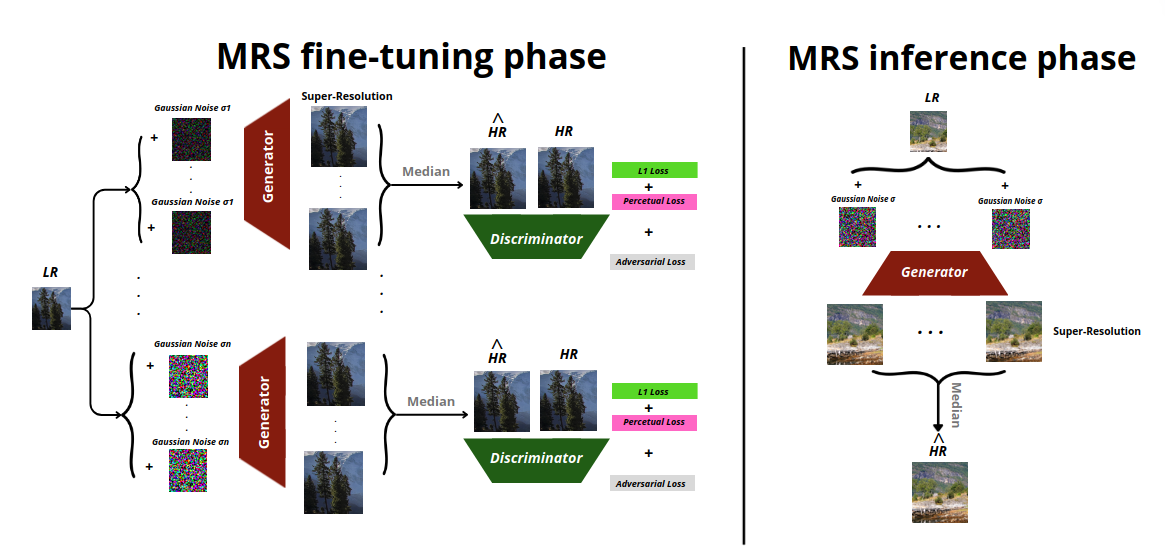}
  \caption{Framework of our proposed CertSR method. In the training part, we add different samples of i.i.d. Gaussians with different standard deviations to the same LR image. We then calculate the median of predictions associated with each standard deviation. In the test part, we use MRS to certify our generator by adding sample i.i.d. Gaussians with the same standard deviation.}
  \label{fig:OurCertSR}
\end{figure*}

\subsection{Median Randomized Smoothing (MRS)}
The MRS is a scalable approach to obtain certified robustness guarantees for any super-resolution neural network. The main principle of this method is to create from one LR image a sample of images by adding Gaussian noise with a certain standard deviation. Then, we get the median of all the predictions pixel-by-pixel. Consequently, we obtain a smoothed model that is certified in an interval of percentiles depending on the perturbation that exists in the input image of the model. More precisely, let  $G\sim N(0,\sigma^{2} I)$,  a Gaussian random variable. The percentile smoothing of a DNN $g_{\theta}: \mathbb{R}^{n}\to \mathbb{R}$ is defined as follows  $$\overline{q}_{p}(x) = \inf\{y\in \mathbb{R} | \mathbb{P}(g_{\theta}(x+G)\leq y)\geq p\},$$ 
$$\underline{q}_{p}(x) = \sup\{y\in \mathbb{R} | \mathbb{P}(g_{\theta}(x+G)\leq y)\leq p\}.$$
We denote $q_{p}(x)$ as the percentile-smoothed function when either definition is applicable. When $p=0.5$ these percentiles are equivalent to the median $q_{0.5}(x)$. Therefore, from \cite{chiang2020detection} we have the following Lemma:
\begin{lemma}\label{certificationlemma}
A percentile-smoothed function $q_{p}$ with adversarial perturbation $\delta$ can be bounded as follows
\begin{equation}
\underline{q}_{\underline{p}}(x)\leq q_{p}(x+\delta)\leq \overline{q}_{\overline{p}}(x), \; \; \;  \forall\Vert\delta\Vert_{2}<\epsilon
\end{equation}
such that  $\overline{p} = \Phi(\Phi^{-1}(p)+\frac{\epsilon}{\sigma})$ and $\underline{p} = \Phi(\Phi^{-1}(p)-\frac{\epsilon}{\sigma})$, where $\Phi$ is the standard Gaussian CDF. 
\end{lemma} 
Here, we are interested in the case $p=0.5$. In this case, the median is bounded between the percentile of  $\overline{p} = \Phi(\frac{\epsilon}{\sigma})$ and $\overline{p} = \Phi(-\frac{\epsilon}{\sigma})$. On the one hand, we observe from Lemma \ref{certificationlemma} that a smaller distance between $\underline{q}_{\underline{p}}(x)$ and $\overline{q}_{\overline{p}}(x)$ indicates a more robust and well-certified model. On the other hand, the bounds of the interval depend on the value of $\frac{\epsilon}{\sigma}$ where $\epsilon$ represents the size of the perturbation against which we aim to certify. Therefore, the choice of $\sigma$ depends on the adversarial attack and the perturbation that exists on LR images. Fortunately, at the inference phase, there is some flexibility in choosing the standard deviation of the Gaussian noise, $\sigma$, that will help us to get a robust and certified SR model.
\subsection{Median Randomized Smoothing for SR}
To create our RSR model which we call CertSR (Certified Super-Resolution) model, we need to go through three essential steps. First, we implement an initial SR model based on a Generative Adversarial Network (GAN) previously trained on clean LR images. Second, we need to fine-tune the SR model on noisy LR images using samples of i.i.d. Gaussians with a specified number of draws and standard deviations. This type of data augmentation will make the SR model more robust to noisy samples. We call this second step $MRS_{Fine-tuning}$. Finally, in a third step that we call $MRS_{Inference}$ phase, we use the median random smoothing method to certify the fine-tuned SR model with a sample of i.i.d. Gaussians associated to a standard deviation (see Figure \ref{fig:OurCertSR}).

\paragraph{Super-Resolution Model}
This study is based on the ESRGAN model \cite{LimBee}, which is a generative adversarial network (GAN) used for super-resolving images. The generator adopts the Residual-in-Residual Dense Block (RRDB) \cite{LedigPhoto-realistic-single-image} structure to improve the quality of the enhanced image. The resolution of the generated images will be enlarged by a factor of 4. We recall that several loss functions are applied during the training. Firstly, the $L_{1}$ loss is used to evaluate the pixel distance between the ground truth (GT) and the super-resolved image. Secondly, the perceptual loss $L_{perc}$ \cite{JustinPerceptuallosses} utilizes the activation features of the pre-trained VGG-19 \cite{SimonyanVery-deep-convolutional} between the GT and the super-resolved image. This loss helps enhance the visual effect of low-frequency components. The third loss is the adversarial loss $L_{adv}$, employed to enhance the texture details of the super-resolved image and make it more realistic. The total loss function is the sum of these three losses:
$$L_{total}= L_{1}+ L_{perc}+ L_{adv}.$$
The Discriminator is structured on a VGG-128 architecture \cite{SimonyanVery-deep-convolutional} and operates under the same principle as the Relativistic GAN \cite{Jolicoeur-Martineau-The-relativistic-discriminator}. It estimates the probability that a real image appears more realistic than a fake one.
\paragraph{CertSR}
We use the pre-trained network generator of the ESRGAN model \cite{LimBee}. Subsequently, we propose to fine-tune this model, denoted $MRS_{fine-tuning}$, on LR images by adding samples of Gaussians noise. We use different standard deviations and for each of them, we choose the same amount of draws\footnote{Note that in our experiments we observed that it is suitable to also use the original image (without adding Gaussian noise).}. Then, we calculate the median of predictions associated with each standard deviation, following the procedure outlined in the fine-tuning phase of Figure \ref{fig:OurCertSR}. Finally, in the inference phase, denoted  $MRS_{inference}$, we use the MRS to certify the SR model with a specific standard deviation, which is a hyperparameter that must be selected to best suit each perturbation, as shown on the right of Figure \ref{fig:OurCertSR}. We emphasize that thanks to the small invariance of the pixel-wise loss on the super-resolved images, at this stage we draw only 21 Gaussian samples in all our experiments to certify our model, which allows us to control this invariability. Moreover, we rely on the LPIPS metric in this context to ensure that we have chosen the best standard deviation.

\section{Experimental Results}\label{Experiments}
In this section, we describe the experimental settings, including the utilized datasets and model configurations.
\subsection{Evaluation Metrics}
We evaluate the performance of different methods by calculating metrics such as Peak-Signal-to-Noise Ratio (PSNR), \cite{YanSingle-imagesuper-resolution}, Structural Similarity Index Measure (SSIM), \cite{WangImagequalityassessment}, and Learned Perceptual Image Patch Similarity (LPIPS), \cite{ZhangTheunreasonableeffectiveness}. PSNR and SSIM are widely used to evaluate image restoration and focus primarily on image fidelity rather than visual quality. LPIPS, on the other hand, places greater emphasis on assessing the similarity of visual features between images. To do this, it uses a pre-trained AlexNet \cite{KrizhevskyImagenetclassification} to extract image features, then calculates the distance between these features. As a result, a lower LPIPS value indicates a closer resemblance between GT and the generated image.

\subsection{Dataset} 

\paragraph{Fine-tuning dataset} We fine-tune the SR models on the DIV2K dataset \cite{AgustssonNtire2017challenge, TimofteNtire2017challenge}  which is a reference commonly used in traditional SISR. Its training set consists of 800 2K resolution images and their respective LR versions, generated by a bicubic downscaling process. These images incorporate no artificial perturbation. We crop the images into 480 × 480 sub-images for our experiments. A scaling factor of 4 was used between the HR images and the 120 × 120 LR images.

\paragraph{Inference dataset}
We assess the performance of our CertSR method on both the clean and the corrupted DIV2K validation dataset \cite{AgustssonNtire2017challenge, TimofteNtire2017challenge}, which contains 100 validation images. Specifically, we corrupt the validation dataset with sensor noise, which is simulated by adding pixel-wise independent Gaussian noise with a mean of 0 and a standard deviation of $0.03$. We also corrupt this dataset by degrading LR images into blurry images. This operation is modeled by smoothing the images with the Gaussian kernel with 10 in size and a standard deviation of $0.3$. Subsequently, we attack the inference dataset with the adversarial attacks defined in section \ref{AdvAttTrain}.

It is also crucial to evaluate our main method on real-world datasets containing various types of synthetic corruptions and sensor noise in LR images. Specifically, we evaluate our method using validation datasets from the NTIRE 2020 Real-World Image Super-Resolution Challenge, Track 1 \cite{LugmayrAndreasNtire2020challenge}, and the AIM 2019 Real World Super-Resolution Challenge, Track 2 \cite{LugmayrAndreasAim2019challenge}. The validation sets comprise artificially degraded versions of the 100 LR images in the DIV2K validation set, together with their corresponding GT. For simplicity, we abbreviate NTIRE 2020 and AIM 2019 as NTIRE and AIM, respectively.

\subsection{Implementation details}

\paragraph{Fine-tuning} is based on the pre-trained ESRGAN \cite{LimBee}. We perform all the fine-tuning methods that we need on a node composed of 8 GPU A100 80Gb with 1.5 Terabytes of RAM and dual AMD processors. We use an Adam optimizer \cite{Adam} with $\beta_{1} = 0.9$ and $\beta_{2} = 0.99$ for both the generator and discriminator with an initial learning rate of $10^{-4}$. For the classical fine-tuning of ESRGAN, as well as for adversarial fine-tuning, we choose 18k iterations and 16 images per batch. Regarding the hyperparameters for adversarial learning, the choices are as follows: (i) Adversarial Learning with FGSM (AD-L-FGSM) has $\epsilon=9/255$. (ii) Adversarial Learning with BIM (AD-L-BIM) uses the same $\epsilon$ as AD-L-FGSM with 2 iterations. (iii) Adversarial Learning with CW (AD-L-CW) employs $c=10^{-2}$, 4 iterations, and utilizes Adam optimization for resolving \ref{CW_Optimization} with a learning rate of $10^{-2}$. (iv) Adversarial Learning with PGD (AD-L-PGD) uses the pre-trained model from \cite{CastilloAngela}. Subsequently, for the $MRS_{Fine-tuning}$ step we take 59k as a number of iterations with 5 images per batch. During this phase, we duplicate the batch training set five times. For the first two batches, we add i.i.d. Gaussian samples with a standard deviation of $\sigma=0.03$. For the next two batches, we add i.i.d. Gaussian samples with a standard deviation of $\sigma=0.2$. The last remaining batch remains unchanged to ensure CertSR considers cleaned images as well. For more details on the hyperparameters of adversarial learning and the $MRS_{Fine-tuning}$ step, please refer to the Appendices \ref{sec:Hyperparam AL} and \ref{sec:Hyperparam MRS}.

\paragraph{Comparison with State-of-the-Art}
We compare our main method CertSR\footnote{See Appendix \ref{sec:Ablation study} for the ablation study.} with other state-of-the-art methods to establish a universal robust baseline for SISR models. For this, we evaluate our results on both clean and corrupted images. We compare our results with ESRGAN \cite{LimBee}, and AD-L-PGD \cite{CastilloAngela}. To ensure a fair comparison, we fine-tune ESRGAN on the DIV2K training set. For real-world images, we also compare our results with the top-performing models on the NTIRE and AIM datasets: Impressionism \cite{JiReal-worldsuper-resolutionviakernelestimationandnoiseinjection} and ESRGAN-FS \cite{FritscheManuelseparationforreal-worldsuper-resolution}, respectively. We use pre-trained weights for Impressionism on NTIRE and DPED \cite{ignatov2017dslr} datasets and for ESRGAN-FS on AIM and DPED datasets. Moreover, we fine-tune Impressionism on AIM and ESRGAN-FS on NTIRE, by employing default parameters from their work.
\begin{table}[ht]
\centering
\setlength{\cmidrulekern}{0.26em} 
\begin{adjustbox}{width=0.47\textwidth}
\begin{tabular}{c|ccc|ccc|ccc|ccc|}
\toprule
  Data    &     \multicolumn{3}{c}{Clean } & \multicolumn{3}{c}{Noisy } & \multicolumn{3}{c}{Blurry}\\
\midrule 
 Method  & PSNR$\uparrow$  & SSIM$\uparrow$ & LPIPS$\downarrow$ & PSNR$\uparrow$  & SSIM$\uparrow$ & LPIPS$\downarrow$ & PSNR$\uparrow$  & SSIM$\uparrow$ & LPIPS$\downarrow$ \\ \hline
ESRGAN \cite{LimBee}  & 27.48  & 0.75  & \color{blue}{0.12} & 20.25  & 0.29  & 0.67 & \color{blue}{22.23}  & \color{blue}{0.62}  & 0.48 \\
  AD-L-PGD \cite{CastilloAngela}  & 26.60  & 0.71  & 0.22 & 22.63  & 0.47  & 0.37 & 22.15  & 0.60  & 0.50\\
 AD-L-FGSM (ours) & 26.28 & 0.70 & 0.34 & 24.84  & 0.57  & 0.32 & 21.95  & 0.59  & 0.53\\
 AD-L-BIM (ours)  & 26.21  & 0.68  & 0.25 & \color{blue}{25.11}  & \color{blue}{0.60}  & \color{blue}{0.29} & 21.93  & 0.58  & \color{blue}{0.48}\\
 AD-L-CW (ours)  & \color{red}{28.41}  & \color{red}{0.77}  & 0.14 & 19.47  & 0.25  & 0.78 & \color{red}{22.34}  & \color{red}{0.62}  & 0.50\\
 \textbf{CertSR (ours)}  & \color{blue}{28.24}  & \color{blue}{0.76}  & \color{red}{0.12} & \color{red}{26.35}  & \color{red}{0.70}  & \color{red}{0.19} & 22.11  & 0.60  & \color{red}{0.44} \\
\bottomrule
\end{tabular}%
\end{adjustbox}
\caption{This table reports the quantitative results of robust and non-robust methods for clean, sensor noise (noisy), and blurry DIV2K validation dataset. In all the tables of this document, the arrows indicate if high $\uparrow$ or low $\downarrow$ values are desired. The best scores are displayed in \textcolor{red}{Red} and the second in \textcolor{blue}{Blue}. }%
\label{tab:tableclean}
\end{table}
\subsection{Evaluation on Clean and Corrupted Images}
In Table \ref{tab:tableclean}, we present a comparison of PSNR, SSIM, and LPIPS values for our CertSR method, the non-robust SR model, ESRGAN , and various RSR models. In the quantitative experiments, we focus on the LPIPS measure, as it has the best correlation with image similarity.  We see from Table \ref{tab:tableclean} that our CertSR method performs well on all three inference datasets. It is important to note that on the clean and noisy dataset, we do not need to use $MRS_{inference}$, using only the $MRS_{fine-tuning}$ we achieve the same results. Furthermore, since the $MRS_{fine-tuning}$ includes both clean and noisy data simultaneously. We obtained a LPIPS value that is almost the same as that of ESRGAN. However, the LPIPS metric value of the ESRGAN model on the noisy dataset is the lowest. Concerning the blurry case, we use the $MRS_{inference}$ on this validation dataset with $\sigma=0.05$. Moreover, we observe that the performance of our CertSR method surpasses that of all other RSR methods. Regarding the other robust models, we can see that AD-L-CW is the best RSR on the clean validation dataset, while AD-L-BIM performs better on the noisy and blurry datasets. Finally, we note that AD-L-FGSM performs better on noisy images than on clean images, which is attributed to the training conducted on attacked images.
\begin{figure}[htp!]  
  \centering 
  \includegraphics[width=\linewidth]{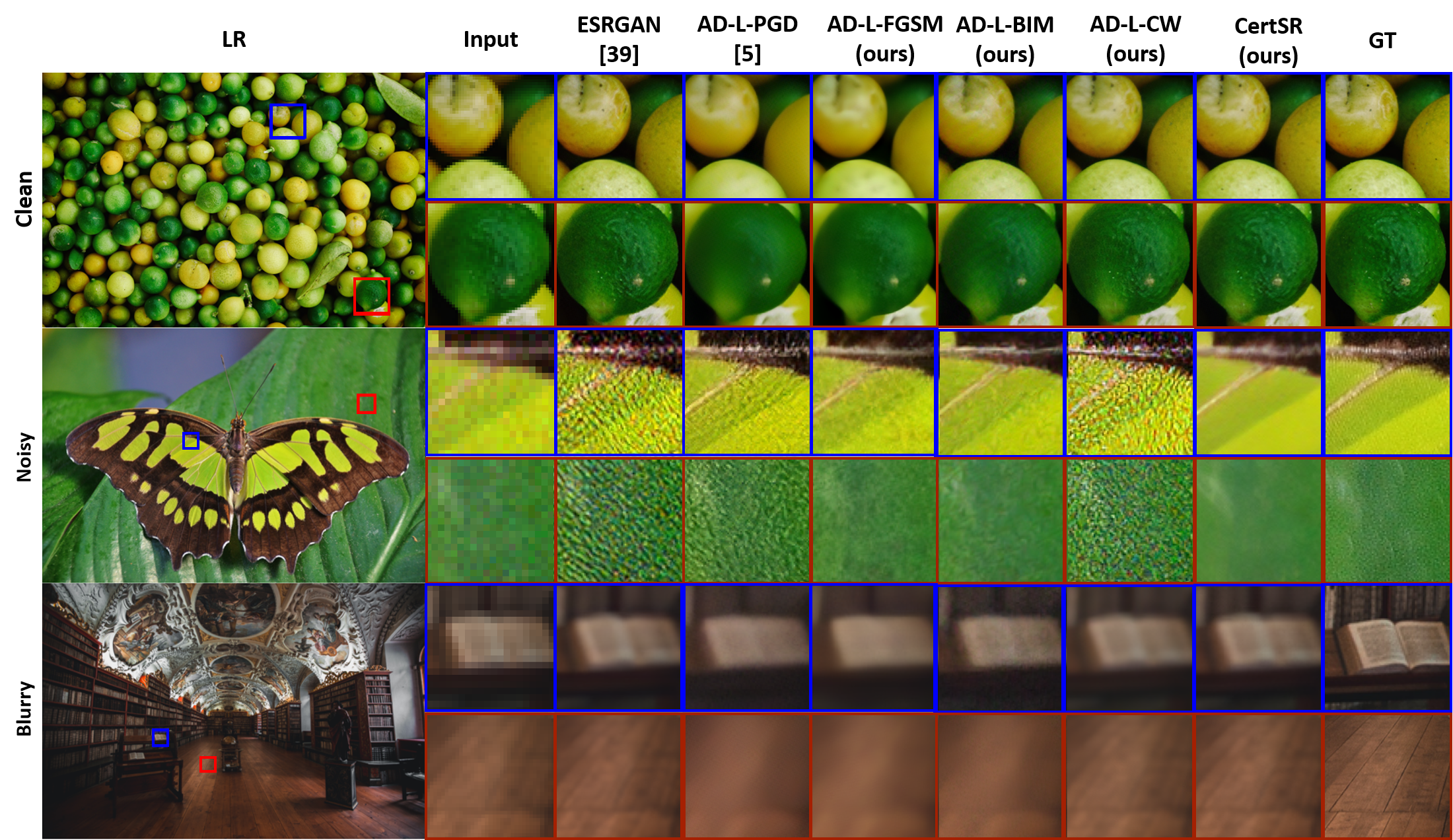}
  \caption{This figure presents the qualitative results of robust and non-robust methods for clean, sensor noise (noisy), and blurry DIV2K validation dataset.}
  \label{fig:figclean}
\end{figure}

Figure \ref{fig:figclean} represents the qualitative results of robust and non-robust methods with respect to the clean, sensor noise, and blurry DIV2K validation dataset. Our CertSR method provides clearer images with richer texture detail and without artifacts, showing that our method is the most robust against noisy and blurry perturbations. On the other hand, we observe that AD-L-PDG and AD-L-FGSM generate very smooth images, and AD-L-BIM introduces some little artifacts in the case where LR images are clean. 
\begin{table}[htp!]
\centering
\setlength{\tabcolsep}{0.5pt}
\setlength{\arrayrulewidth}{0.3mm}
\setlength{\cmidrulekern}{0.26em} 
\begin{adjustbox}{width=0.47\textwidth}
\begin{tabular}{c|ccc|ccc|ccc|ccc|ccc|}
\toprule
  Adversarial attacks    &     \multicolumn{3}{c}{FGSM } & \multicolumn{3}{c}{BIM } & \multicolumn{3}{c}{PGD } & \multicolumn{3}{c}{CW }\\
\midrule
 Method  & PSNR$\uparrow$  & SSIM$\uparrow$ & LPIPS$\downarrow$ & PSNR$\uparrow$  & SSIM$\uparrow$ & LPIPS$\downarrow$ & PSNR$\uparrow$  & SSIM$\uparrow$ & LPIPS$\downarrow$  & PSNR$\uparrow$  & SSIM$\uparrow$ & LPIPS$\downarrow$ \\ \hline
 ESRGAN \cite{LimBee}  &  16.70  & 0.18  & 0.70 & 14.97  & 0.15  & 0.76 & 17.83  & 0.19  & 0.83 & 16.43 & 0.23  & 0.69\\
 AD-L-PGD \cite{CastilloAngela}  & 21.74  & 0.50  & 0.36 & 19.45  & 0.45  & 0.44 & 24.21  & \color{blue}{0.60}  & \color{blue}{0.24} & 25.15 & \color{blue}{0.67}  & \color{blue}{0.24}\\
 AD-L-FGSM (ours)  & \color{red}{25.55}  & \color{red}{0.70}  & \color{red}{0.19}  & 23.48  & \color{blue}{0.60}  & 0.29 & 21.56  & 0.39  & 0.46 & 24.13  & 0.64  & 0.32\\
 AD-L-BIM (ours)  & 24.17  & 0.60  & 0.27 & \color{blue}{23.79}  & 0.59  & \color{blue}{0.26} & \color{blue}{24.65}  & 0.59  & 0.33 & \color{blue}{25.57}  & 0.65  & 0.25\\
 AD-L-CW (ours)  & 4.72  & 0.23  & 0.99 & 12.83  & 0.09  & 0.91 & 15.39  & 0.13  & 0.95 & 18.37  & 0.33  & 0.61\\
 \textbf{CertSR (ours)}  & \color{blue}{24.72}  & \color{blue}{0.64}  & \color{blue}{0.27} & \color{red}{24.28}  & \color{red}{0.64}  & \color{red}{0.25} & \color{red}{25.09}  & \color{red}{0.67}  & \color{red}{0.24} & \color{red}{26.66}  & \color{red}{0.72}  & \color{red}{0.18} \\
\bottomrule
\end{tabular}%
\end{adjustbox} 
\caption{This table shows the quantitative results concerning robust and non-robust methods against the most relevant adversarial attacks. The best scores are displayed in \textcolor{red}{Red} and in \textcolor{blue}{Blue}.}%
\label{tab:tabagainstattacks}
\end{table}

Table \ref{tab:tabagainstattacks} presents the quantitative results of the robust and non-robust methods against the adversarial attacks. To study this, we place ourselves in the worst-case scenario, which means we test the universality of our CertSR's robustness against the same attacks that were used to build RSR models. It is important to mention that in the validation part, we use $MRS_{inference}$ against each adversarial attack with respect to different standard deviations. More precisely, against PGD (see \ref{PGDAttack}) and FGSM (see \ref{FGSMAttack}) attacks, we certify our model with $\sigma = 0.06$. Against the BIM attack (see \ref{BIMAttack}), we choose $\sigma = 0.07$, and against the CW attack (see \ref{CWAttack}), we use $\sigma = 0.03$ (please consult the Appendix \ref{sec:Hyperparam MRS} to see how these hyperparameters have been selected).  Therefore, we see from Table \ref{tab:tabagainstattacks} that our main method achieves the best performance against all adversarial attacks with respect to PSNR, SSIM and LPIPS metrics, except against ADV-L-FGSM, where CertSR is the second-best method against FGSM attacks. Therefore, we can say that CertSR is the most globally robust SR method against adversarial attacks.
\begin{figure}[htp!]  
  \centering 
  \includegraphics[width=\linewidth]{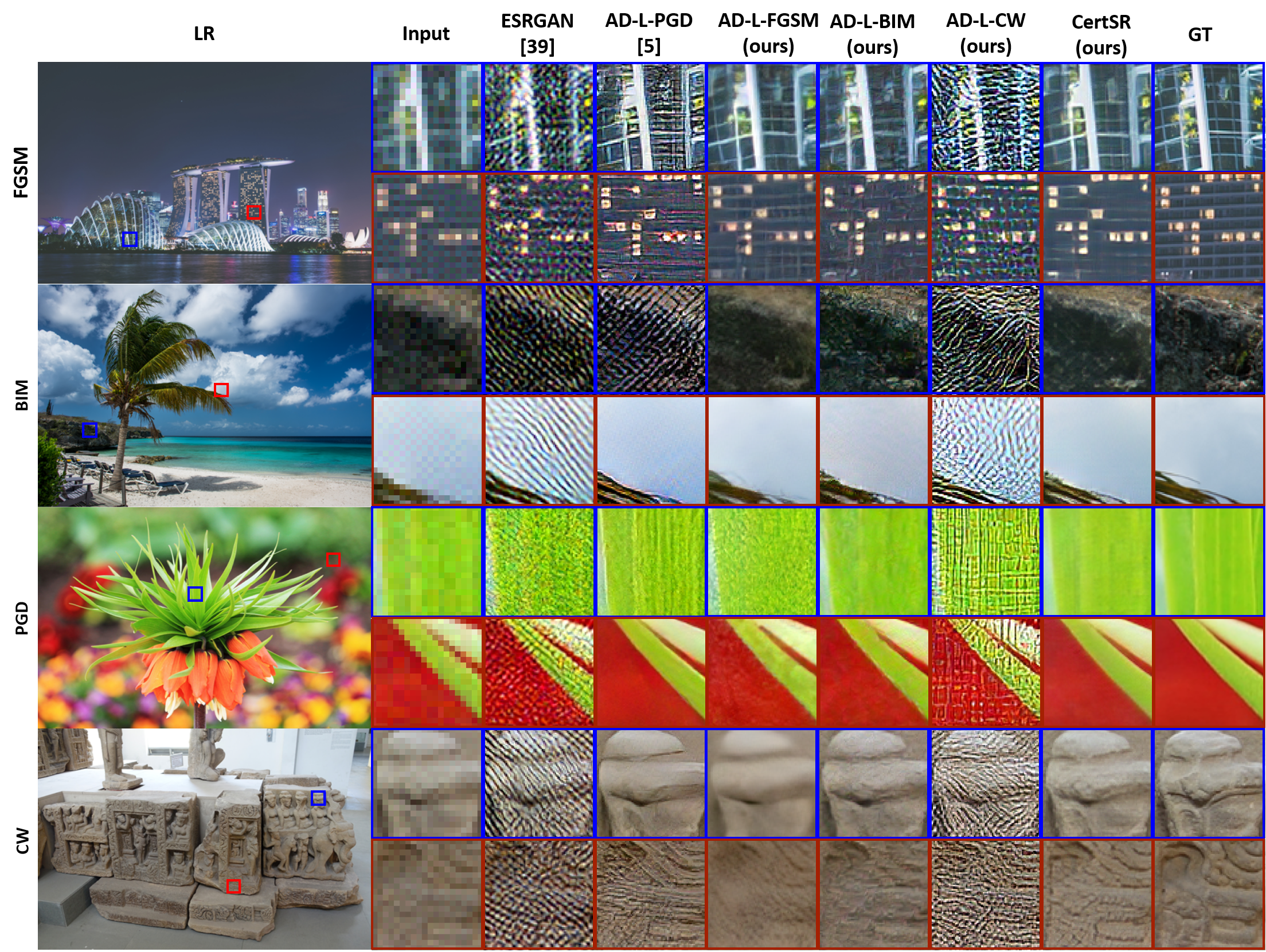}
  \caption{This figure provides qualitative results concerning robust and non-robust methods against the most relevant adversarial attacks.}
  \label{fig:figAdveAttacks}
\end{figure}

In Figure \ref{fig:figAdveAttacks}, we present qualitative results concerning CertSR's robustness against the most relevant adversarial attacks. Visually, it is clear that CertSR produces super-resolved images that are superior to those of other RSR models. The images generated by these RSR models show noticeable artifacts. This figure illustrates that even models trained with a specific adversarial attack remain somewhat vulnerable when subjected to a similar attack. We observe that the weakest robust SR model is AD-L-CW. This is related to the fact that even CW attack has the advantage of being the optimal and strongest attack, it also has the disadvantage of being the most difficult to learn. 

\begin{table*}[htp!]
\setlength{\tabcolsep}{1.1pt}
\setlength{\arrayrulewidth}{0.1mm}
\setlength{\cmidrulekern}{0.22em} 
\makebox[\textwidth][c]{ 
\begin{tabular}{|c|c|c||ccc||ccc||ccc||ccc|}
\toprule
    \multirow{2}{*}{Method} & \multirow{2}{*}{Training Data} & \multirow{2}{*}{Fine-tuning Data} & \multicolumn{3}{c}{PSNR$\uparrow$ } & \multicolumn{3}{c}{SSIM$\uparrow$} & \multicolumn{3}{c}{LPIPS$\downarrow$ } \\ & & 
  & NTIRE  & AIM & Avg & NTIRE  & AIM & Avg & NTIRE  & AIM & Avg \\ 
  \hline
  Bicubic &  &  & 25.51 & \color{blue}{22.35} & 23.93 & 0.67 & \color{blue}{0.62} & 0.65 & 0.63 & 0.68 & 0.66\\
  \hline
   \multirow{3}{*}{ESRGAN-FS \cite{FritscheManuelseparationforreal-worldsuper-resolution}} &  & NTIRE & 24.59 & 22.07 & 23.33 & \color{blue}{0.69} & \color{red}{0.63} & \color{red}{0.66} & 0.25 & 0.47 & 0.36\\
 & Flickr2K & AIM & 19.56 & 20.82 & 20.19 & 0.31 & 0.51 & 0.41 & 0.56 & 0.39 & 0.48 \\
 &  & DPEP & 17.79 & 20.15 & 18.97 & 0.34 & 0.53 & 0.43 & 0.51 & 0.47 & 0.49 \\
 \hline
\multirow{3}{*}{Impressionism \cite{JiReal-worldsuper-resolutionviakernelestimationandnoiseinjection}} &  & NTIRE & 24.82 & 21.47 & 23.15 & 0.66 & 0.54 & 0.60 & \color{blue}{0.23} & 0.52 & 0.37\\
 & Flickr2K & AIM & 19.65 & 21.89 & 20.77 & 0.29 & 0.60 & 0.45 & 0.67 & 0.41 & 0.54 \\
 &  & DPEP & 17.53 & 18.84 & 18.18 & 0.34 & 0.49 & 0.41 & 0.60 & 0.47 & 0.53 \\
 \hline
 ESRGAN \cite{LimBee} & Flickr2K & DIV2k  & 21.94  & 21.95  & 21.03  & 0.39  & 0.55 & 0.49  & 0.56  & 0.51 & 0.53 \\
 \hline
  AD-L-PGD \cite{CastilloAngela} & Flickr2K & DIV2K  & 24.31  & 21.99 & 23.15  & 0.65  & 0.60 & 0.62  & 0.23  & 0.37 & \color{blue}{0.30}\\
 \hline 
 AD-L-FGSM (ours) & Flickr2K & DIV2k  & \color{blue}{25.55}  & \color{red}{22.70}  & \color{blue}{24.20}  & 0.65  & \color{red}{0.63} & 0.64  & 0.30  & 0.42 & 0.36 \\
 \hline
 AD-L-BIM (ours) & Flickr2K & DIV2K  & 25.35  & 22.31 & 23.95  & 0.63  & 0.59 & 0.61  & 0.26  & \color{blue}{0.36} & 0.31 \\
 \hline
 AD-L-CW (ours) & Flickr2K  & DIV2K  & 21.25  & 21.86 & 21.63  & 0.37  & 0.58 & 0.48  & 0.63  & 0.47 & 0.55 \\
 \hline
 \textbf{CertSR (ours)} & Flickr2K & DIV2K  &  \color{red}{26.67} & 21.75 & \color{red}{24.21}  & \color{red}{0.71}  & 0.59 & \color{blue}{0.65}  & \color{red}{0.21}  & \color{red}{0.33} & \color{red}{0.27}  \\
\bottomrule
\end{tabular}%
}
\caption{\textbf{Quantitative results on Real-World Images.} We present the quantitative results of reference metrics between our method, state-of-the-art methods, and robust and non-robust models on NTIRE and AIM validation datasets. \textcolor{red}{Red} and \textcolor{blue}{Blue} colors highlight the best two scores.
\textbf{Bold} represents the best method for LPIPS metric for both datasets.}%
\label{tab:tabagainstRealWorld}
\end{table*}

\subsection{Evaluation on Real-World Images}
\begin{figure*}[htp!]  
  \centering 
  \includegraphics[width=\linewidth]{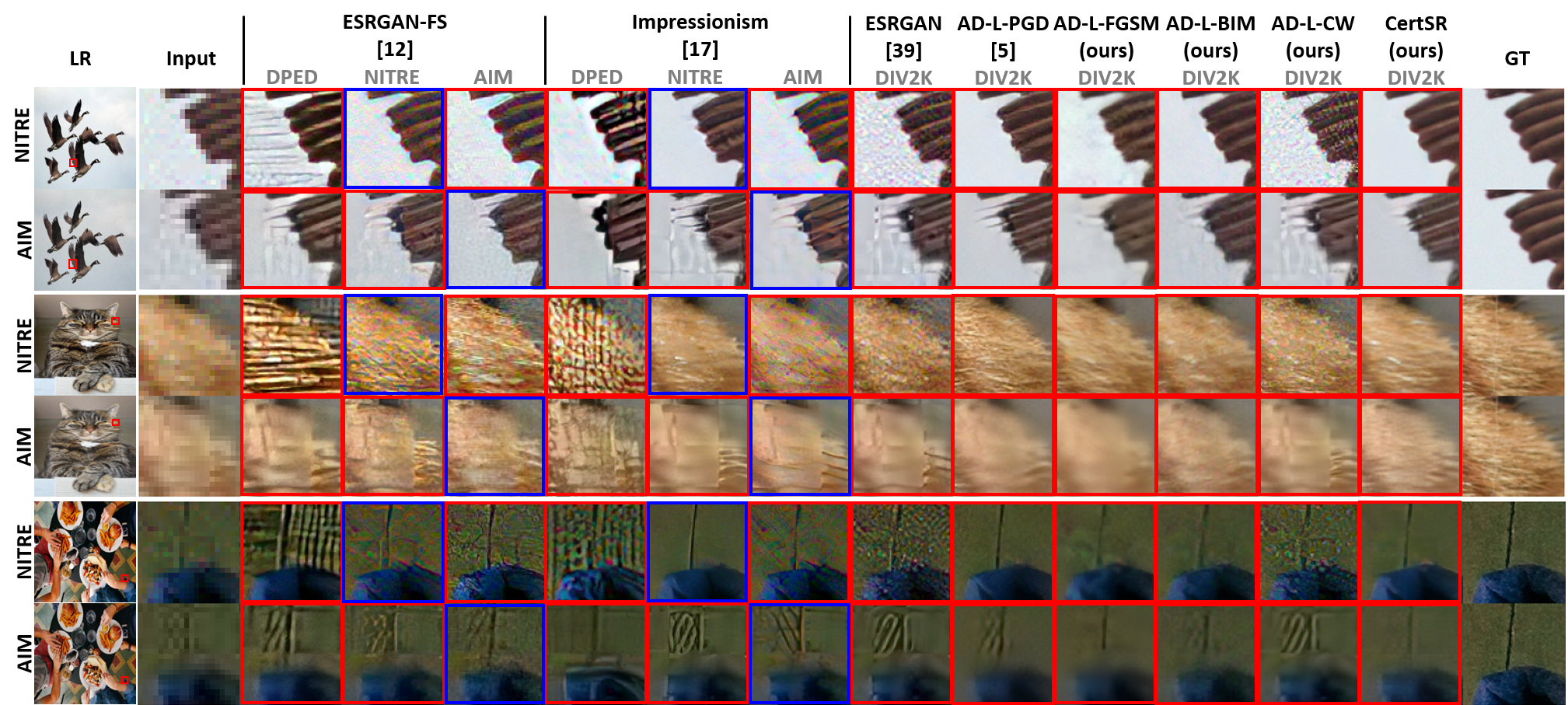}
  \caption{\textbf{Qualitative results on Real-World Images.} Comparison between the proposed methods including CertSR and state-of-the-art RSR method (AD-L-PGD \cite{CastilloAngela}), for two corruption datasets: NTIRE and AIM. For reference, we show the input, the results of ESRGAN-FS method \cite{FritscheManuelseparationforreal-worldsuper-resolution}, Impressionism method \cite{JiReal-worldsuper-resolutionviakernelestimationandnoiseinjection} and the ground-truth (GT). \textcolor{blue}{Blue} frames denote training and validation on the same dataset. \textcolor{red}{Red} frames denote training and validation on different datasets. The training dataset is indicated in gray just below the name of the methods.}
\label{fig:figureRWImages}
\end{figure*}

Table \ref{tab:tabagainstRealWorld} presents the quantitative results of reference metrics for CertSR method, state-of-the-art methods and RSR models on both the NTIRE and AIM validation datasets. We observe that CertSR achieves the best LPIPS performance without any training or fine-tuning on these datasets. AD-L-CW and ESRGAN achieve the worst LPIPS on both validation datasets. We also observe that AD-L-BIM is more performant than AD-L-PGD on the AIM. These results are visually confirmed in Figure \ref{fig:figureRWImages}. For the $MRS_{inference}$ phase, we choose $\sigma=0.03$ and $\sigma=0.06$ for NTIRE and AIM respectively. Please refer to the Appendix \ref{sec:Hyperparam MRS} to see how these hyperparameters have been selected.

It is important to note that, we also test the proposed CertSR method on other SR models besides ESRGAN, on both NTIRE and AIM validation datasets, to demonstrate that the method can enhance the accuracy and robustness of other initial SR models. See the Appendix \ref{sec:Model Agnostic} for more details.  
\section{Conclusion}
In this work, we explore the fruitful relationship between Robust Super-Resolution (RSR) and real-world SR. Our main finding is the demonstration that the most universal model in terms of robustness to different adversarial attacks is also the more robust to unseen natural noise in the LR input real-world images. This important insight is based on a study conducted on two different types of RSR models: one type built from various adversarial training techniques (including the existing RSR model using PGD attack \cite{CastilloAngela} and new RSR models that we built from FGSM, BIM and the CW attacks) and another original one built from a certification technique that leverages MRS procedure with Gaussian noise. Our experiments on synthetic and real datasets show that, compared to the RSR models AD-L-PGD \cite{CastilloAngela} AD-L-FGSM, AD-L-BIM, AD-L-CW, the proposed model CertSR, is the most universal in terms of robustness to adversarial attacks and is also the one that achieves the best results on real-world SR. We also show that the CertSR achieved state-of-the-art results in particular with the LPIPS metric. We expect that this finding will encourage further study of the RSR approach to tackle noise in real-world SR.
\paragraph{Acknowledgements} This publication was made possible by the use of the FactoryIA supercomputer, financially supported by the Ile-de-France Regional Council. The authors thank Patrick Hede for his technical support in using FactoryIA.

{
    \small
    \bibliographystyle{ieeenat_fullname}
    \bibliography{main}
}

\newpage

\appendix

\section{Ablation study}
\label{sec:Ablation study}
In this section, we conduct an ablation study to investigate the performance of the proposed CertSR method by removing each of the two main components to understand their contribution to the overall method. Specifically, we explore the effects of both the Median Randomized Smoothing (MRS) fine-tuning phase and the MRS inference phase (see Figure 2 in the main paper) and compare them with the global method that includes both (CertSR). In Table \ref{table:AblationStudy}, we report the results of this study. We observe that both MRS components have a slight positive impact on the SR model. However, together these two components give much better results, leading us to the proposed method, CertSR. 

We note that "ESRGAN" indicates the fine-tuning of ESRGAN \cite{LimBee} on the DIV2K dataset \cite{AgustssonNtire2017challenge}. The "ESRGAN+MRS$_{FT}$" method involves fine-tuning the ESRGAN model using only Median Randomized Smoothing (MRS), while "ESRGAN+MRS$_{Inf}$" indicates the use directly of MRS in the inference phase of ESRGAN. Finally, CertSR is a combination of "ESRGAN+MRS$_{FT}$" and "ESRGAN+MRS$_{Inf}$".

\begin{table}[htbp!]
\centering
\begin{adjustbox}{max width=0.46\textwidth}  
\begin{tabular}{cccccc}
\toprule
\multirow{2}{*}{Dataset} & \multirow{2}{*}{Metrics} & \multicolumn{4}{c}{SR Methods}\\
\cmidrule{3-6}
 & & ESRGAN & ESRGAN+MRS$_{FT}$ &  ESRGAN +MRS$_{Inf}$  &  CertSR\\
\midrule
\multirow{3}{*}{AIM} & PSNR $\uparrow$ & \color{red}{21.95} & 21.88 & \color{blue}{21.97} & 21.75  \\
 & SSIM $\uparrow$ & 0.55 & \color{blue}{0.56} & 0.53 & \color{red}{0.59} \\
 & LPIPS $\downarrow$ & 0.51  & \color{blue}{0.47} & 0.48 & \color{red}{0.33} \\
\midrule
\multirow{3}{*}{NTIRE} & PSNR $\uparrow$ & 21.94 & \color{red}{26.90} & 22.16 & \color{blue}{26.67} \\
 & SSIM $\uparrow$ & 0.39  & \color{blue}{0.69} & 0.40 & \color{red}{0.71} \\
 & LPIPS $\downarrow$ & 0.56 &  \color{blue}{0.22} & 0.55 &  \color{red}{0.21} \\
\bottomrule
\end{tabular}
\end{adjustbox}
\caption{\textbf{Ablation study}. We present the comparison of reference
metrics between our method and each of their component independently. Red and blue colors highlight the best two scores.}
\label{table:AblationStudy}
\end{table}
 
Firstly, by examining Table \ref{table:AblationStudy}, we observe an enhancement in the performance of "ESRGAN+MRS$_{FT}$" compared to "ESRGAN." This improvement is attributed to the fine-tuning phase where MRS introduces Gaussian random noise to the input images. This strategy fosters model invariance to small changes in the input, consequently enhancing generalization to previously unseen data. It is important to note, that due to the Gaussian data augmentation utilized in the fine-tuning phase, this method serves as an alternative to regularization in neural networks with the Jacobian of the model \cite{bishop1995training}. This alternative becomes especially valuable for SR tasks where applying Jacobian-based regularization is often impractical due to the substantial dimensions of the input and output. Secondly, we observe that the "ESRGAN+MRS$_{Inf}$" method also improves the performance of ESRGAN, particularly concerning the LPIPS metrics. However, this method is not as effective when applied independently; its efficacy increases notably when used after "ESRGAN+MRS$_{FT}$." This can be attributed to the sensitivity of ESRGAN to Gaussian noise.
\section{CertSR with other SR models}
\label{sec:Model Agnostic}
In this section, we will test our CertSR method on some other SR models. The purpose of this study is to demonstrate that our method can enhance the precision and robustness of any SR model. Moreover, this enhancement comes at no additional cost. For this reason, we choose the SR models EDSR \cite{lim2017enhanced} and NINASR \cite{torchSR}. We will then apply the certification method to them (see Figure 2 in the main paper). We denote CertEDSR and CertNINASR as the models EDSR and NINASR after the certification process, respectively. In Table \ref{table:cert_on_others_models}, we present the results that we obtained after and before the certification method on AIM \cite{LugmayrAndreasAim2019challenge} and NTIRE \cite{AgustssonNtire2017challenge} datasets.

\begin{table}[htbp!]
\centering
\begin{adjustbox}{max width=0.46\textwidth}  
\begin{tabular}{cccccc}
\toprule 
\multirow{2}{*}{Dataset} & \multirow{2}{*}{Metrics} & \multicolumn{4}{c}{SR Methods}\\
\cmidrule{3-6}
 & & EDSR & CertEDSR &  NINASR  &  CertNINASR\\
\midrule
\multirow{3}{*}{AIM} & PSNR $\uparrow$ &  22.57 & 22.32 & 22.22 & 22.24  \\
 & SSIM $\uparrow$ & 0.60 & 0.53 & 0.59 &  0.61\\
 & LPIPS $\downarrow$ & 0.60  & 0.57 & 0.60 & 0.49 \\
\midrule
\multirow{3}{*}{NTIRE} & PSNR $\uparrow$ & 25.57 & 26.67 & 24.79 & 27.61 \\
 & SSIM $\uparrow$ & 0.64  & 0.70 & 0.63 &  0.74 \\
 & LPIPS $\downarrow$ & 0.57 &  0.47 & 0.57 &  0.37 \\
\bottomrule
\end{tabular}
\end{adjustbox}
\caption{We show a comparison of reference metrics between two  SR models before and after applying the certification method that we propose.}
\label{table:cert_on_others_models}
\end{table}

In this study, similarly to ESRGAN,  we fine-tune both the EDSR and NINASR models on the DIV2K training dataset. This involves applying MRS$_{FT}$ to both models with identical standard deviations, $\sigma_{1}= 0.03$ and $\sigma_{1}= 0.2$, corresponding to the Gaussians samples. Next,  we apply MRS$_{inf}$ to both models. To be specific, we draw 21 i.i.d Gaussians samples with a standard deviation of $\sigma=0.1$ to derive CertEDSR and CertNINASR results on the AIM dataset. Regarding the results on the NTIRE datasets, we maintain the same number of draws and we use $\sigma=0.005$.

\section{Comparison with RSR via regularization}

In this section, we will regularize the ESRGAN neural network with the gradient of the loss function, a well-known method to ensure the stability of the neural network against input corruption and perturbation. In addition, this method allows for penalizing large changes in the output neural network model, enforcing a smoothness prior. This method has been employed in several works focused on classification tasks, as seen in, for instance, \cite{drucker1991double, sokolic2017robust, varga2017gradient}.

We recall that the loss function used to train or to fine-tune the ESRGAN is given by $$L_{total}= L_{1, perc}+ L_{adv}.$$
where, $L_{1, perc}= L_{1}+ L_{perc}$. Here, $L_{1}$ loss is the pixel distance, $L_{perc}$ is the perceptual loss, and $L_{adv}$ is the adversarial loss. Due to the gradient regularization that we will apply, the new total loss function becomes as follows: 
\begin{equation}\label{regularization}
\tag{$s_{1}$}
L_{reg}= L_{total}+ \lambda *\Vert \nabla_{x} L_{1, perc}\Vert,    
\end{equation}
where $\lambda$ is a hyperparameter. It is important to point out that the method we use in this part is similar to the regularization used in \cite{ChoiJun-Ho2}. Besides, we regularize with the gradient of $L_{1}$ and $L_{perc}$ because our aim is to get a robust SR model both pixel-wise and perceptually.
\begin{table}[htbp!]
\centering
\begin{adjustbox}{max width=0.46\textwidth}  
\begin{tabular}{cccccc}
\toprule
\multirow{2}{*}{Dataset} & \multirow{2}{*}{Metrics} & \multicolumn{4}{c}{SR Methods}\\
\cmidrule{3-6}
 & & ESRGAN & AD-L-PGD  &  ESRGAN-Reg  &  CertSR\\
\midrule
\multirow{3}{*}{AIM} & PSNR $\uparrow$ & 21.91 & \color{red}{21.99} & \color{blue}{21.97} & 21.75  \\
 & SSIM $\uparrow$ & 0.55 & \color{red}{0.60} & 0.55 & \color{blue}{0.59} \\
 & LPIPS $\downarrow$ & 0.51  &  \color{blue}{0.37} & 0.50 & \color{red}{0.33} \\
\midrule
\multirow{3}{*}{NTIRE} & PSNR $\uparrow$ & 21.94 & \color{blue}{24.31} & 21.69 & \color{red}{26.67} \\
 & SSIM $\uparrow$ & 0.39  & \color{blue}{0.65} & 0.38 & \color{red}{0.71} \\
 & LPIPS $\downarrow$ & 0.56 &  \color{blue}{0.23}  & 0.57 &  \color{red}{0.21} \\
\bottomrule
\end{tabular}
\end{adjustbox}
\caption{We present the comparison of reference
metrics between RSR via gradient regularization, RSR via adversarial learning with PGD attack, ESRGAN and our CertSR}
\label{table:RegMethod}
\end{table}

The result given from this study is shown in Table \ref{table:RegMethod}, where we compare this method of regularization, denoted as ESRGAN-Reg, with other methods such as ADV-L-PGD \cite{CastilloAngela}, constructed via adversarial learning using the PGD attack, ESRGAN fine-tuned in DIV2K, and our CertSR. We note that in our experiment, the best hyperparameter that yielded good results is $\lambda=0.001$. On the other hand, from Table \ref{table:RegMethod}, we can deduce that this method of robustness is not very efficient in the SR task, notably for real-world SR.

\section{Hyperparametrs for Median Randomized Smoothing (MRS)}
\label{sec:Hyperparam MRS}
In this section, we explore the impact of the hyperparameters for the proposed MRS fine-tuning and MRS inference, as shown in Figure 2 in the main paper. 
\subsection{Hyperparametrs for MRS fine-tuning}
The MRS fine-tuning method has been done on DIV2K training dataset. However, for the validation of this method, we did it in AIM and NTIRE validation dataset. We would like to emphasize that in this phase, we chose two types of Gaussian samples, with each sample corresponding to a standard deviation. Additionally, for each Gaussian sample, we drew it two times randomly. In Table \ref{table:MRS-fine-tuning} we show the impact of the hyperparameters $\sigma_1$ and $\sigma_2$ on the performance of the MRS fine-tuning phase, validated on the AIM and NTIRE validation datasets based on LPIPS metric.
\begin{table}[htbp!]
\centering
\begin{adjustbox}{max width=0.5\textwidth}
\begin{tabular}{ccccccccc}
\toprule
\multirow{2}{*}{Dataset} & \multirow{2}{*}{Metric} & \multirow{2}{*}{ Std $\sigma_{1}$} & \multicolumn{5}{c}{Std $\sigma_{2}$}\\
\cmidrule{4-9}
 & & & 0.01 & 0.02 & 0.03 & 0.04 & 0.05 & 0.06\\
\midrule
\multirow{4}{*}{AIM} 
 & \multirow{4}{*}{LPIPS} & 0.1 & 0.48 & 0.48 & 0.48 & 0.48 & 0.48 & 0.48 \\
 & & 0.2 & 0.49 & 0.48 & \textbf{0.47}  & 0.47 & 0.48 & 0.48 \\
 & & 0.3 & 0.48 & 0.48 & 0.49 & 0.48 & 0.48 & 0.48  \\
 & & 0.4 & 0.49 & 0.49 & 0.48 & 0.47 & 0.48 & 0.48 \\
 & & 0.5 & 0.49 & 0.48 & 0.49 & 0.49 & 0.48 & 0.48 \\
 & & 0.6 & 0.48 & 0.48 & 0.48 & 0.48 & 0.48 & 0.48 \\
\midrule
\multirow{4}{*}{NTIRE} 
 & \multirow{4}{*}{LPIPS} & 0.1 & 0.30 & 0.26 & 0.24 & 0.23 & 0.25 & 0.25\\
 & & 0.2 & 0.33 & 0.26 &  \textbf{0.22} & 0.24 & 0.24 & 0.25\\
 & & 0.3 & 0.36 & 0.27 & 0.22 & 0.24 & 0.24 & 0.26\\
 & & 0.4 & 0.37 & 0.28 & 0.24 & 0.26 & 0.27 & 0.28\\
 & & 0.5 & 0.40 & 0.25 & 0.22 & 0.24 & 0.26 & 0.30\\
 & & 0.6 & 0.40 & 0.27 & 0.23 & 0.26 & 0.27 & 0.29 \\
\bottomrule 
\end{tabular}
\end{adjustbox}
\caption{We report the impact of the hyperparameters $\sigma_1$ and $\sigma_2$ on the performance of the MRS fine-tuning phase, validated on the AIM and NTIRE validation datasets.}
\label{table:MRS-fine-tuning}
\end{table}

\subsection{Hyperparametrs for MRS Inference}
After the MRS fine-tuning, We represent the performance of the MRS inference against the adversarial attacks on the DIV2K validation dataset and the real-world validation datasets. 

In Table \ref{table:InferenceMRSAdv}, we show the impact of the hyperparameter $\sigma$ on the performance of MRS$_{inf}$ validated on the AIM and NTIRE validation datasets based on PSNR, SSIM, and LPIPS metrics. We point out that the number of draws used in the inference phase is the same, which is 21.
\begin{table}[htbp!]
\centering
\begin{adjustbox}{max width=0.48\textwidth}  
\begin{tabular}{ccccccccccc}
\toprule
\multirow{2}{*}{attack} & \multirow{2}{*}{Metrics} & \multicolumn{8}{c}{Hyperparameter $\sigma$} \\
\cmidrule{3-11}
 & & $0.005$  & $0.01$ & $0.02$  & $0.03$ & $0.04$ & $0.05$ & $0.06$ & $0.07$ &$0.08$ \\
\midrule
\multirow{3}{*}{FGSM} & PSNR & 19.73 & 19.92 & 20.73 & 21.74 & 22.95 & 24.11 & 24.72 & 24.92 & 24.92 \\
 & SSIM & 0.35 & 0.36 & 0.40 & 0.46 & 0.53 & 0.60  & 0.64 & 0.65 & 0.65 \\
 & LPIPS & 0.48 & 0.48 & 0.44 & 0.39  & 0.34 & 0.29 & \textbf{0.27} & 0.28 & 0.30\\
\midrule
\multirow{3}{*}{BIM} & PSNR & 17.38 & 17.61 & 18.60  & 19.72 & 20.10 & 22.35 & 23.53 & 24.28 & 24.60 \\
 & SSIM & 0.28 & 0.29 & 0.33 & 0.38 & 0.45 & 0.53 & 0.60 & 0.64 & 0.65 \\
 & LPIPS & 0.56 & 0.55 & 0.51 & 0.47 & 0.41 & 0.33 & 0.27 & \textbf{0.25} & 0.27 \\
\midrule
\multirow{3}{*}{PGD} & PSNR & 22.15 & 22.68 & 23.91  & 24.42 & 24.62 & 24.85 & 25.09 & 25.19 & 25.15 \\
 & SSIM & 0.47 & 0.51 & 0.60 & 0.64 & 0.65 & 0.66 & 0.67 & 0.67 & 0.68 \\
 & LPIPS & 0.50 & 0.46 & 0.38 & 0.32 & 0.28 & 0.25 & \textbf{0.24} & 0.25 & 0.28 \\
 \midrule
\multirow{3}{*}{CW} & PSNR & 21.69 & 24.87 & 26.46  & \textbf{26.66} & 26.48 & 26.25 & 26.09 & 25.94 & 25.73 \\
 & SSIM & 0.48 & 0.58 & 0.65 & 0.71 & 0.72 & 0.71 & 0.70 & 0.69 & 0.68 \\
 & LPIPS & 0.38 & 0.22 & 0.19 & \textbf{0.18} & 0.18 & 0.19 & 0.21 & 0.24 & 0.27 \\
\bottomrule
\end{tabular}
\end{adjustbox}
\caption{We present the performance of the MRS inference phase, on attacked DIV2K validation dataset.}
\label{table:InferenceMRSAdv}
\end{table}

In Table \ref{table:InferenceMRS}, we present the impact of the hyperparameter $\sigma$ on the performance of $MRS_{inf}$ validated on the AIM and NTIRE validation datasets based on PSNR, SSIM, and LPIPS metrics. The number of draws used in the inference phase is also 21.

\begin{table}[htbp!]
\centering
\begin{adjustbox}{max width=0.48\textwidth}  
\begin{tabular}{ccccccccccc}
\toprule
\multirow{2}{*}{Dataset} & \multirow{2}{*}{Metrics} & \multicolumn{8}{c}{Hyperparameter $\sigma$} \\
\cmidrule{3-11}
 & & $0.005$  & $0.01$ & $0.02$  & $0.03$ & $0.04$ & $0.05$ & $0.06$ & $0.07$ &$0.08$ \\
\midrule
\multirow{3}{*}{AIM} & PSNR & 21.91 & 22.07 & 22.17 & 22.07 & 21.90 & 21.77 & 21.75 & 21.98 & 22.01 \\
 & SSIM & 0.57 & 0.60 & 0.61 & 0.61 & 0.60 & 0.59  & 0.59 & 0.60 & 0.60 \\
 & LPIPS & 0.46 & 0.45 & 0.42 & 0.38  & 0.36 & 0.34 & \textbf{0.33} & 0.34 & 0.36\\
\midrule
\multirow{3}{*}{NTIRE} & PSNR & 26.86 & 26.93 & 27.02  & 26.67 & 26.41 & 26.17 & 26.29 & 26.15 & 25.80 \\
 & SSIM & 0.69 & 0.70 & 0.71 & \textbf{0.71} & 0.70 & 0.69 & 0.68 & 0.68 & 0.69 \\
 & LPIPS & 0.23 & 0.22 & 0.22 & \textbf{0.21} & 0.21 & 0.22 & 0.24 & 0.27 & 0.28 \\
\bottomrule
\end{tabular}
\end{adjustbox}
\caption{We report the impact of the hyperparameters $\sigma$ on the performance of the MRS inference phase, based on reference metrics validated on the AIM and NTIRE validation datasets.}
\label{table:InferenceMRS}
\end{table}

\section{Hyperparameters for adversarial Learning}
\label{sec:Hyperparam AL}
In this section, we explore the impact of the hyperparameters for the proposed adversarial learning methods based on adversarial attacks (FGSM, BIM, and CW) that we use to build RSR models. 
\subsection{Adversarial Learning with FGSM (AD-L-FGSM)}
In Table \ref{table:HyperparametersFGSM}, we present the results of the AD-L-FGSM model for different values of the hyperparameter of the FGSM adversarial attack, which is $\epsilon$,  representing the step size for the allowed perturbation. We report results on the AIM and NTIRE datasets for different metrics, namely PSNR, SSIM, and LPIPS.
\begin{table}[htbp!]
\centering
\begin{adjustbox}{max width=0.45\textwidth}  
\begin{tabular}{ccccccc}
\toprule
\multirow{2}{*}{Dataset} & \multirow{2}{*}{Metrics} & \multicolumn{5}{c}{Hyperparameter $\epsilon$} \\
\cmidrule{3-7}
 & & $1/255$  & $3/255$ & $6/255$  & $9/255$ & $10/255$ \\
\midrule
\multirow{3}{*}{AIM} & PSNR & 22.18 & 22.59 & 22.64 & 22.70 & 22.77 \\
 & SSIM & 0.56 & 0.60 & 0.62 & 0.63 & 0.62  \\
 & LPIPS & 0.44 & 0.42 & 0.43 & 0.42 & 0.46\\
\midrule
\multirow{3}{*}{NTIRE} & PSNR & 22.98 & 23.50 & 24.66 & 25.55 & 25.50 \\
 & SSIM & 0.46 & 0.49 & 0.57 & 0.65 & 0.64   \\
 & LPIPS & 0.46 & 0.44 & 0.35 & 0.30 & 0.32\\
 \midrule
\end{tabular}
\end{adjustbox}
\caption{We present the performance of the AD-L-FGSM model for different values of the hyperparameter $\epsilon$ on the AIM and NTIRE validation datasets with respect to reference metrics.}
\label{table:HyperparametersFGSM}
\end{table}

\subsection{Adversarial Learning with BIM (AD-L-BIM)}

In Table \ref{table:BIMHyperparameter}, we present the results of the AD-L-BIM model for different values of the hyperparameters of the BIM adversarial attack. The hyperparameters of this attack are composed of $\alpha$, which represent the step of the perturbations are  and $T$ the number of iterations. We report the results on the AIM and NTIRE datasets with respect to different metrics, namely PSNR, SSIM, and LPIPS.
\begin{table}[htbp!]
\centering
\begin{adjustbox}{max width=0.43\textwidth}
\begin{tabular}{cccccccc}
\toprule
\multirow{2}{*}{Dataset} & \multirow{2}{*}{Metrics} & \multirow{2}{*}{Iteration $T$} & \multicolumn{5}{c}{Hyperparameter $\alpha$}\\
\cmidrule{4-8}
 & & & 1/255 & 3/255 & 6/255 & 9/255 & 10/255 \\
\midrule
\multirow{13}{*}{AIM} 
 & \multirow{4}{*}{PSNR} & 2 & 22.36 & 18.16 & 16.87 & 22.31 & 17.93 \\
 & & 3 & 22.71 & 17.89 & 17.51 & 17.64 & 18.03 \\
 & & 4 & 22.26 & 16.75 & 18.11 & 17.85 & 17.29 \\
 & & 5 & 17.57 & 16.32 & 16.44 & 18.19 & 19.05 \\
 \cmidrule{2-8}
 & \multirow{4}{*}{SSIM} & 2 & 0.61 & 0.29 & 0.29 & 0.59 & 0.29 \\
 & & 3 & 0.62 & 0.39 & 0.30 & 0.28 & 0.35 \\
 & & 4 & 0.60 & 0.22 & 0.32 & 0.29 & 0.27 \\
 & & 5 & 0.30 & 0.22 & 0.21 & 0.30 & 0.40 \\
 \cmidrule{2-8}
 & \multirow{4}{*}{LPIPS} & 2 & 0.46 & 0.68 & 0.76 & 0.36 & 0.73 \\
 & & 3 & 0.45 & 0.76 & 0.80 & 0.86 & 0.79 \\
 & & 4 & 0.47 & 0.75 & 0.70 & 0.74 & 0.82  \\
 & & 5 & 0.86 & 0.87 & 0.72 & 0.71 & 0.63 \\
\midrule
\multirow{13}{*}{NTIRE} 
 & \multirow{4}{*}{PSNR} & 2 & 25.53 & 18.37 & 17.02 & 25.35 & 18.62\\
 & & 3 & 25.62 & 23.55 & 17.84 & 18.31 &18.29 \\
 & & 4 & 25.56 & 18.49 & 18.59 & 18.05 &17.79\\
 & & 5 & 17.77 & 24.06  & 16.83 & 18.93 & 20.03\\
 \cmidrule{2-8}
 & \multirow{4}{*}{SSIM} & 2 & 0.64 & 0.23 & 0.24 & 0.63 & 0.28\\
 & & 3 & 0.65 & 0.48 &  0.25 & 0.27 & 0.30\\
 & & 4 & 0.64 & 0.28 & 0.28 & 0.25 & 0.26 \\
 & & 5 & 0.27 & 0.51 & 0.20 & 0.27 & 0.40 \\
 \cmidrule{2-8}
 & \multirow{4}{*}{LPIPS} & 2 & 0.34 & 0.69 & 0.76 & 0.26 & 0.72\\
 & & 3 & 0.33 & 0.41 &  0.77 & 0.83 &0.80\\
 & & 4 & 0.33 & 0.76 & 0.71 & 0.74 &0.80\\
 & & 5 & 0.85 & 0.40 & 0.71 & 0.70 & 0.61\\
\bottomrule
\end{tabular}
\end{adjustbox}
\caption{We present the performance of the AD-L-BIM model for different values of the hyperparameters $\alpha$ (the step of the adversarial attack) and $T$ (number of iterations) on the AIM and NTIRE validation datasets with respect to reference metrics.}
\label{table:BIMHyperparameter}
\end{table}

\subsection{Adversarial Learning with CW (AD-L-CW)}

In Table \ref{table:CWHyperparametrs}, we present the results of the AD-L-CW model for different values of the hyperparameters of the CW adversarial attack. The hyperparameters of this attack are composed of $c$, which controls the trade-off between the L2 norm of the perturbation and $T$ the number of iterations to minimize the following problem:
\begin{equation}\label{s2}
\tag{$s_{2}$}
\min_{\delta}(\Vert\delta \Vert_{2} -c\cdot\mathcal{ L}(f_{\theta}(x), y)), \text{ such that } x+\delta \in [0,1]^{n}.  
\end{equation} We report the results on the AIM and NTIRE datasets with respect to different metrics, namely PSNR, SSIM, and LPIPS.
\begin{table}[htbp!]
\centering
\begin{adjustbox}{max width=0.40\textwidth}
\begin{tabular}{ccccc}
\toprule
\multirow{2}{*}{Dataset} & \multirow{2}{*}{Metrics} & \multirow{2}{*}{Iterations $T$} & \multicolumn{2}{c}{Hyperparameter $c$}\\
\cmidrule{4-5}
 & & & $10^{-2}$ & $1$ \\
\midrule
\multirow{16}{*}{AIM} 
 & \multirow{5}{*}{PSNR} & 1 & 21.51 & 4.60 \\
 & & 2 & 5.35 & 4.59 \\
 & & 3 & 4.64 & 4.58 \\
 & & 4 & 21.86 & 5.21 \\
 & & 5 & 4.72 & 5.37 \\
 \cmidrule{2-5}
 & \multirow{5}{*}{SSIM} & 1 & 0.52 & 0.11\\
 & & 2 & 0.12 & 0.02 \\
 & & 3 & 0.01 & 0.23 \\
 & & 4 & 0.58 & 0.06 \\
 & & 5 & 0.22 & 0.07 \\
 \cmidrule{2-5}
 & \multirow{5}{*}{LPIPS} & 1 & 0.51 & 1.01\\
 & & 2 & 1.06 & 0.91 \\
 & & 3 & 1.09 & 1.16 \\
 & & 4 & 0.47 & 1.13 \\
 & & 5 & 0.99 & 1.06 \\
\midrule
\multirow{16}{*}{NTIRE} 
 & \multirow{5}{*}{PSNR} & 1 & 20.87 & 4.60 \\
 & & 2 & 5.27 & 4.59 \\
 & & 3 & 4.65 & 4.57 \\
 & & 4 & 21.25 & 4.99 \\
 & & 5 & 4.72 & 5.00 \\
 \cmidrule{2-5}
 & \multirow{5}{*}{SSIM} & 1 & 0.32 & 0.11\\
 & & 2 & 0.12 & 0.01 \\
 & & 3 & 0.03 & 0.06 \\
 & & 4 & 0.37 & 0.01 \\
 & & 5 & 0.24 & 0.01 \\
 \cmidrule{2-5}
 & \multirow{4}{*}{LPIPS} & 1 & 0.67 & 1.01\\
 & & 2 & 1.06 & 0.91 \\
 & & 3 & 1.16 & 1.28 \\
 & & 4 & 0.63 & 1.30 \\
 & & 5 & 0.99 & 1.22 \\
\bottomrule
\end{tabular}
\end{adjustbox}
\caption{We present the performance of the AD-L-CW model for different values of the hyperparameters $c$  (controls the trade-off between the L2 norm of the perturbation) and $T$ (number of iterations to minimize \ref{s2}) on the AIM and NTIRE validation datasets with respect to reference metrics.}
\label{table:CWHyperparametrs}
\end{table}


\end{document}